\documentclass[twocolumn,prl]{revtex4}

\usepackage{graphicx}

\def\prl{Phys. Rev. Lett. }

\def\be{\begin{equation}}
\def\ee{\end{equation}}
\def\ba{\begin{eqnarray}}
\def\ea{\end{eqnarray}}

\begin{document}

\title{Doping dependent isotope effects of the quasi-1D electron-phonon system:\\
comparison with the high-temperature superconductors}

\author{I. P. Bindloss}
\affiliation{Department of Physics, University of California,  
Los Angeles, California 90095--1547}
\date{\today}

\begin{abstract}
The weak-coupling quantum phase diagrams of the one-dimensional (1D) Holstein-Hubbard and 
Peierls-Hubbard models are computed near half-filling, using a multi-step renormalization group 
technique.  If strong enough, the electron-phonon interaction induces a spin gap.  The spin gap, which 
determines the superconducting pairing energy, depends strongly on the band filling and decreases 
monotonically as the system is doped away from half-filling.  However, the superconducting 
susceptibility exhibits a different doping dependence; it can vary non-monotonically with doping and 
exhibit a maximum at an "optimal" value of the doping.  For a quasi-1D array of weakly coupled, 
fluctuating 1D chains, the superconducting transition temperature $T_c$ exhibits a similar 
non-monotonic doping dependence.  The effect of changing the ion mass (isotope effect) on $T_c$ is 
found to be largest near half-filling and to decrease rapidly upon doping away from half-filling.  The 
isotope effect on the spin gap is the opposite sign as the isotope effect on $T_c$.  We discuss 
qualitative similarities between these results and properties of the high-temperature superconductors.
\end{abstract}

\maketitle

Recent experiments in the high-temperature superconductors suggest the presence
of a strong, ubiquitous, yet unconventional electron-phonon (el-ph) interaction \cite{lanzara,chung}.
It has been known for some time that these materials exhibit large and strongly doping
dependent oxygen isotope effects \cite{isotope,pseudogap,pendepth}
that cannot be described with the conventional BCS theory used
for ordinary metals.
In this Letter, we show that the quasi-1D electron gas
coupled to phonons exhibits highly unconventional doping dependent
isotope effects that are qualitatively similar
to those observed in the high-temperature superconductors.

The best understood non-Fermi liquid
is the spin-charge separated one-dimensional electron gas (1DEG) \cite{1DEG}.  The properties
of the 1DEG coupled to phonons are dramatically different from a conventional
metal coupled to phonons \cite{zimanyi,voit,ian,spectral}.  Unlike in a Fermi liquid,
in 1D the el-ph interaction is strongly
renormalized, and the renormalization is strongly
affected by direct electron-electron (el-el) interactions.
Due to these renormalization effects, a weak, retarded el-ph interaction
is capable of inducing a spin gap and causing a divergent
superconducting susceptibility, even when the 
el-el repulsion is the dominant microscopic interaction \cite{ian}.  Unlike the case in 3D,
in 1D this can occur without a large amount of retardation.  In contrast
to BCS theory, the pairing energy
and superconducting susceptibility are very sensitive to the band filling.

In the cuprates, both the superconducting transition temperature $T_c$ and
the isotope effect exponent $\alpha_{T_c} = - d \ln T_c / d \ln M$,
which describes changes in $T_c$ induced by changes
in the oxygen mass $M$, exhibit highly unconventional ({\it i.e.} non-BCS)
doping dependencies.  In BCS theory, $T_c$ is
only weakly dependent on the carrier concentration,
and $\alpha_{T_c}$ has the universal value of $1/2$.
In the cuprates, $T_c$ exhibits a maximum as a function of doping, and
the isotope effect is not universal--it is
strongly doping and somewhat material dependent.
For the underdoped cuprates, $\alpha_{T_c}\approx 1$ (indicating that, at least in this region,
phonons play an important role in the superconductivity).
As the doping increases, $\alpha_{T_c}$ decreases, usually
dropping below 0.1 near optimal doping \cite{isotope}.  The origin
of this behavior remains a mystery of
high-$T_c$ superconductivity that any successful
microscopic theory of pairing should explain.
Below we provide a microscopic theory that is capable
of rationalizing it.

In this Letter, we
compute $\alpha_{T_c}$ for a quasi-1DEG coupled to phonons, under the
assumption that charge density wave (CDW) order is dephased by spatial
or dynamic fluctuations of the 1D chains \cite{zachar,Nature}.
For many choices of the parameters, $\alpha_{T_c}$ is larger than the BCS value
at small dopings, then drops below the BCS value as the doping is increased --
the same behavior observed in the cuprate superconductors. 
We show that the quasi-1DEG coupled to phonons displays a
strongly doping dependent
$T_c$ that can exhibit a maximum as a function of doping.
This behavior occurs despite the fact that the pairing energy, determined by the spin gap $\Delta_s$,
is a monotonically decreasing function of increasing doping.
We also compute the isotope exponent $\alpha_{\Delta_s} = - d \ln \Delta_s / d \ln M$.  We find 
$\alpha_{\Delta_s} < 0$,
the same sign as the isotope effect observed on the pseudogap temperature
in the cuprates \cite{pseudogap}.

The technique we employ is the multi-step renormalization group (MSRG) method
described in detail in a previous paper \cite{ian}.
This method treats
el-el and el-ph interactions on equal footing and properly treats
the quantum phonon fluctuations. In it,
we start with a microscopic el-ph model
and integrate out high energy degrees of freedom, via an RG transformation.
This is done in multiple steps, as elaborated below.
At low energies one obtains an effective
field theory that is the same as the ordinary 1DEG,
except for phonon induced renormalizations of the el-el interactions and bandwidth.
The accuracy of this analytic technique in computing weak-coupling phase diagrams
was demonstrated in Ref. \onlinecite{ian}
by comparison with exact numerical results.  We shall apply it to two models of
interacting, spinful 1D electrons coupled to phonons:

The 1D Peierls-Hubbard (Pei-Hub) Hamiltonian is
\ba
&& \nonumber {\cal H}_{\rm Pei-Hub} = \nonumber -\sum_{i,\sigma}\left[t - \gamma 
(u_{i+1}-u_{i})\right](c_{i,\sigma}^\dagger c_{i+1,\sigma} + {\rm H.C.})\\
&& \nonumber \;\;\;\; + \, \sum_i \left[ \frac{p_i^2}{2M} + \frac{\kappa}{2} (u_{i+1}-u_i)^2 \right] +  
U \sum_i n_{i,\uparrow} n_{i,\downarrow} \, .
\label{P}
\ea
In this model, acoustic phonons couple to electrons by modifying the bare hopping matrix element $t$ by 
the
el-ph coupling strength $\gamma$ times the relative displacements $u_{i+1} - u_i$
of two neighboring ions \cite{SSH}.
The last term is the Hubbard interaction.
For this model, we shall approximate
the phonon dispersion by its value at the zone boundary of $2 \sqrt{\kappa/M} \equiv \omega_0$,
since the el-ph interaction vanishes at zero momentum transfer.

The 1D Holstein-Hubbard (Hol-Hub) Hamiltonian is
\ba
\nonumber {\cal H}_{\rm Hol-Hub} = \nonumber -t\sum_{i,\sigma}(c_{i,\sigma}^\dagger c_{i+1,\sigma} + 
{\rm H.C.}) +  \sum_i \left[ \frac{p_i^2}{2M} + \frac{K}{2} q_i^2 \right]  \\
\nonumber + \; g \sqrt{2 M \omega_0} \, \sum_i q_i n_i + U \sum_i n_{i,\uparrow} n_{i,\downarrow} \: 
.\;\;\;\;\;\;\;\;\;\;\;\;\;\;\;
\label{H}
\ea
Here dispersionless optical
phonons with coordinate $q_i$ and frequency
$\omega_0 = \sqrt{K/M}$
couple to the
electron density $n_i = \sum_{\sigma} c_{i,\sigma}^\dagger c_{i,\sigma}$ with el-ph coupling strength 
$g$ \cite{Holstein}.

It is convenient to define the dimensionless quantities
\ba
\label{couplingconstants}
\nonumber && \lambda_{\rm Pei} = 2 N_0 (\gamma \sin k_F)^2 /\kappa \, , \;\; \lambda_{\rm Hol}  =  N_0 
g^2 / \omega_0 \, ,\\
\nonumber && \, {\bar U} =  U/(\pi v_F) \, , \;\;\,  \delta = \ln(\mu/\omega_0) / \ln(E_F/\omega_0) \, 
,
\ea
where $N_0 \equiv 2/(\pi v_F)$.  As usual, $v_F$, $k_F$, and $E_F$ are the Fermi velocity, momentum, 
and energy, respectively.
(We have set $\hbar$ and the lattice parameter equal to unity.)
$\mu$ is the chemical potential measured with
respect to its value at half-filling.
In this Letter, we study the range $\omega_0 < \mu < E_F$
($0 < \delta < 1$).
The doping concentration relative
to half-filling is given
by $x \approx N_0 \omega_0 (E_F/\omega_0)^\delta$. 
Since the MSRG method is perturbative, it is only
quantitatively accurate for $\lambda_{\nu}, {\bar U} \ll 1$,
but is believed to be qualitatively accurate for
$\lambda_{\nu}, {\bar U} \lesssim 1$ \cite{ian}. (The subscript
$\nu$ stands for Pei or Hol.)  

\begin{figure}
\includegraphics[width=0.48\textwidth]{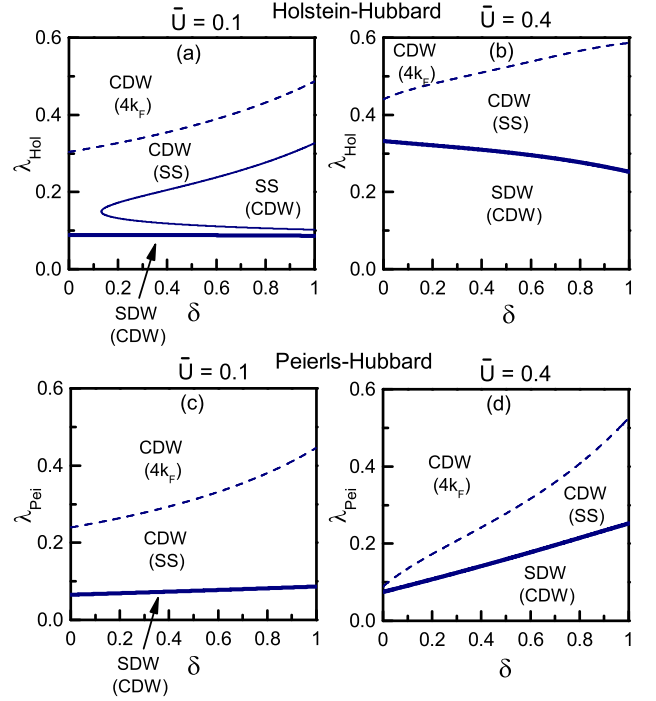}
\caption{\label{fig_1} 
$T = 0$ phase diagrams for the Hol-Hub model (panels a and b),
and Pei-Hub model (panels c and d).  
(a) and (c) are for ${\bar U} = 0.1$; (b) and (d) are for ${\bar U} = 0.4$.
For all diagrams, $E_F/\omega_0 = 5$.
Parenthesis indicate a sub-dominant susceptibility.
SDW stands for $2 k_F$ spin density wave, CDW stands for $2 k_F$ charge density wave,
SS stands for singlet superconductivity, and $4 k_F$ stands for $4 k_F$ charge density wave.
}
\end{figure}

Fig. \ref{fig_1} presents the $\lambda_\nu - \delta$ phase diagrams of the above models,
computed with MSRG for several fixed values of ${\bar U}$. 
The phase boundaries separate regions where various types of order have divergent
susceptibilities in the low temperature limit.  The susceptibility
that diverges most strongly ({\it i.e.} dominates) is shown without parenthesis;
parenthesis indicate a susceptibility that diverges less strongly.  The charge sector is gapless 
everywhere
in the phase diagrams.  Above the thick solid line,
the system is spin-gapped and described as a Luther-Emery liquid (LEL) \cite{LEL}.
Below this line, it is a gapless, quantum-critical Luttinger liquid (LL) \cite{1DEG}.

The thick solid line in Fig. \ref{fig_1} is determined by
integrating out degrees of freedom between $E_F$ and $\omega_0$,
and then requiring that the total effective backward-scattering interaction
$g_1^{\rm tot}(\omega_0) = \bar U/(1 + {\bar U}l_0) - \lambda_1(\omega_0)$
is zero.  Here $l_0 \equiv \ln(E_F/\omega_0)$ and $\lambda_1(\omega_0) > 0$ is the effective
strength of the backward scattering (momentum transfer near $2 k_F$)
portion of the el-ph interaction.
Below the thick line in Fig. \ref{fig_1}, $g_1^{\rm tot}(\omega_0) > 0$
and the RG flows carry the effective $g_1^{\rm tot}$ to zero at low energies,
signifying the stability of the LL fixed point.  Above the thick line, $g_1^{\rm tot}(\omega_0) < 0$
and the RG flows carry $g_1^{\rm tot}$ to minus infinity at low energies,
indicating the existence of a spin gap.

$\lambda_1(\omega_0)$ is determined in two steps:  First, one integrates from $E_F$ to $\mu$
using the RG flow equations that
govern half-filled systems, resulting in an effective $\lambda_1$ of \cite{ian}
\be
\lambda_1(\mu) = \left(\frac{\lambda_{\rm Hol}}{1 - \lambda_{\rm Hol} X / {\bar 
U}}\right)\sqrt{\frac{1-c{\bar U} l_0}{(1+c{\bar U} l_0)^3}}
\ee
or
\be
\lambda_1(\mu) = \left(\frac{\lambda_{\rm Pei}}{1 - \lambda_{\rm Pei} Y / {\bar 
U}}\right)\sqrt{\frac{1}{[1-(c{\bar U} l_0)^2]^3}}
\ee
for the Hol-Hub and Pei-Hub models respectively, where
$X \equiv 4\left[1 - \sqrt{(1 - c{\bar U} l_0)/(1 + c{\bar U} l_0)}\right] - 2\arcsin(c{\bar U} l_0)$, 
$Y \equiv 2c \bar U l_0/\sqrt{1 - (c{\bar U} l_0)^2}$, and
$c \equiv 1 - \delta$.
Next, $\lambda_1(\mu)$ is used as the initial value to integrate
from $\mu$ to $\omega_0$, employing the RG flow equations
that govern incommensurate systems, resulting in \cite{ian}
\be
\lambda_1(\omega_0) = \left(\frac{\lambda_1(\mu)}{1 - \lambda_1(\mu)Z/{\bar U}}\right) 
\sqrt{\frac{\exp(\delta {\bar U}l_0)}{(1 + \delta {\bar U}l_0)^3}}
\ee
for either model, where $Z \equiv \int_0^{\delta {\bar U}l_0} du \sqrt{\exp(u)/(1+u)^3}$.
The condition $g_1^{\rm tot}(\omega_0) = 0$ then determines the following critical
values for the microscopic el-ph couplings:
\ba
\label{g1zeroH}
\lambda_{\rm Hol}^{\rm Gap} &=& {\bar U}\left\{\left[(1 + {\bar U}l_0)S_3 + Z\right]/S_1 \, + \, X 
\right\}^{-1} , \\
\lambda_{\rm Pei}^{\rm Gap} &=& {\bar U}\left\{\left[(1 + {\bar U}l_0)S_3 + Z\right]/S_2 \, + \, Y 
\right\}^{-1} ,
\label{g1zeroP}
\ea
where we defined
$S_1 = (1 + c{\bar U}l_0)^{3/2}(1 - c{\bar U}l_0)^{-1/2}$,
$S_2 = \left[1 - (c{\bar U}l_0)^2\right]^{3/2}$, and
$S_3 = e^{\delta {\bar U}l_0/2}(1 + \delta {\bar U}l_0)^{-3/2}$.
The system is a spin-gapped LEL for
$\lambda_{\nu} > \lambda^{\rm Gap}_{\nu}$.

In the LEL phase, the portion of the
singlet superconductivity (SS)
and $2 k_F$ CDW susceptibility that is
potentially strongly divergent as $T \rightarrow 0$ 
is given by $\chi_{\rm SS} = (\pi v_F)^{-1} (\Delta_s/E_F)(T/E_F)^{1/K_c^{\rm eff} - 2}$
and $\chi_{\rm CDW} = (\pi v_F)^{-1} (\Delta_s/E_F)(T/E_F)^{K_c^{\rm eff} - 2}$
respectively, where the spin gap is
$\Delta_s = \omega_0 e \exp[-1/|g_1^{\rm tot}(\omega_0)|]$ \cite{ian},
and the effective charge Luttinger exponent
after integrating out states between $E_F$ and $\omega_0$ is
$K_c^{\rm eff} = \sqrt{\left[2 + 2g_4^{\rm tot} + g_c^{\rm tot}(\omega_0)\right]/\left[2 + 2g_4^{\rm 
tot} - g_c^{\rm tot}(\omega_0)\right]}$.
Here $g_c^{\rm tot}(\omega_0) = g_c^{\rm el}(\mu) - \lambda_1(\omega_0) + 2 \lambda_2$
and $g_4^{\rm tot} = {\bar U}/2 - \lambda_2$, where the forward scattering el-ph interaction 
$\lambda_2$
is given by $\lambda_2 = \lambda_{\rm Hol}$ for the Hol-Hub model and $\lambda_2 = 0$
for the Pei-Hub model.
The contribution of the Hubbard interaction to $g_c^{\rm tot}(\omega_0)$, given by $g_c^{\rm el}(\mu) = 
-{\bar U}/(1 - c {\bar U}l_0)$,
is obtained by integrating out states between $E_F$ and $\mu$, since this contribution is 
unrenormalized
below $\mu$.  Note that $K_c^{\rm eff}$ is the effective value at low energies because
$g_c^{\rm tot}$ is not further renormalized below $\omega_0$ \cite{footnote1}.  			 

The SS susceptibility is the dominant one if $K_c^{\rm eff} > 1$.
The thin solid line in Fig. \ref{fig_1}a is the critical line determined by $K_c^{\rm eff} = 1$,
given by
\be
\lambda_{\rm Hol}^{\rm{SS}, \pm} = {\bar U}\left[B \pm \sqrt{B^2 - S_1 AC/2}\right] ,
\label{Kc1}
\ee
where $A = (S_1 X + Z)^{-1}$, $B = [(2S_1 - S_3)A + C]/4$, and $C = (1-c{\bar U}l_0)^{-1}$.
$\chi_{\rm SS}$ is dominant if the two conditions $S_1 AC < 2B^2$ and
$\lambda_{\rm Hol}^{\rm{SS}, -} < \lambda_{\rm Hol} < \lambda_{\rm Hol}^{\rm{SS}, +}$
are met.
The thin solid line is absent in Fig. \ref{fig_1}b
because $S_1 AC > 2B^2$ everywhere.
This line is never present
in the Pei-Hub model with ${\bar U} > 0$, because then 
$K_c^{\rm eff} < 1$ always.

\begin{figure}
\includegraphics[width=0.42\textwidth]{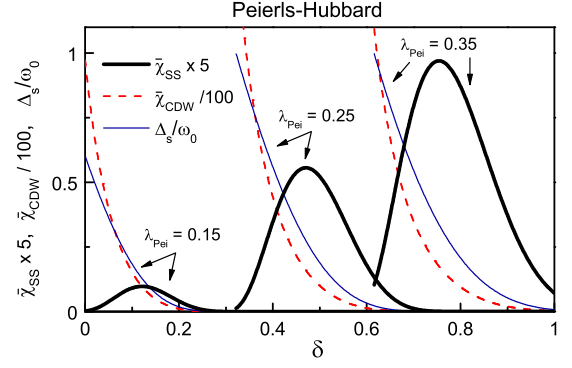}
\caption{\label{fig_2} 
Doping dependence of the dimensionless
singlet superconducting susceptibility ${\bar \chi}_{\rm SS}$ (thick solid lines),
CDW susceptibility ${\bar \chi}_{\rm CDW}$
(dashed lines), and spin gap (thin solid lines), for the Pei-Hub model with ${\bar U} = 0.4$, 
$E_F/\omega_0 = 5$,
and various values of $\lambda_{\rm Pei}$ (labeled in plot).
${\bar \chi}_{\rm SS}$ and ${\bar \chi}_{\rm CDW}$ were computed for $T/\omega_0 = 0.1$.
}
\end{figure}

For $1/2 < K_c^{\rm eff} < 1$, $\chi_{\rm SS}$ is still divergent as $T \rightarrow 0$, but, for a 
single chain
1DEG, $\chi_{\rm CDW}$ is more strongly divergent.  The dashed lines in Fig. \ref{fig_1} are determined 
by $K_c^{\rm eff} = 1/2$, which
leads to the following critical values for the microscopic el-ph interactions 
\ba
\lambda_{\rm Hol}^{\rm CDW} &=& {\bar U}\left[D + \sqrt{D^2 + 5 S_1 AE/4}\right] ,\\
\lambda_{\rm Pei}^{\rm CDW} &=& {\bar U}\left[\left(S_3/E + Z\right)/S_2 + Y\right]^{-1} ,
\label{Kc12}
\ea
where $D = A[S_1(4 - 5 EX) - 5(EZ + S_3)]/8$ and $E = (6/{\bar U} + 3)/5 - C$.
If $\lambda_{\nu} > \lambda^{\rm CDW}_{\nu}$,
then $K_c^{\rm eff} < 1/2$, which means that SS is not divergent.

\begin{figure}
\includegraphics[width=0.445\textwidth]{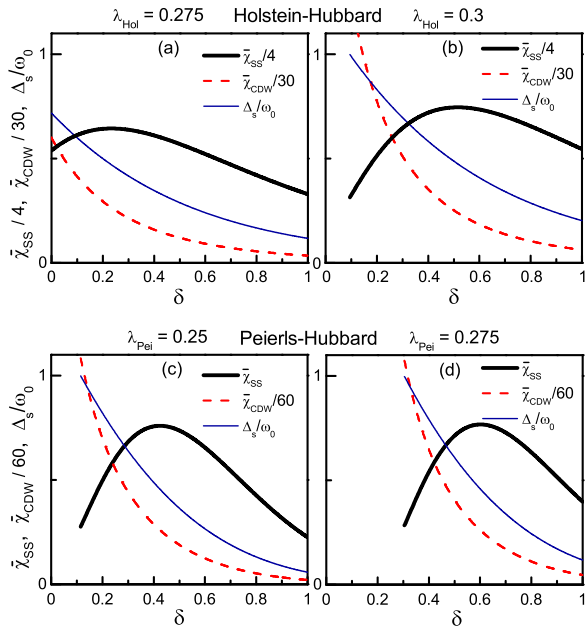}
\caption{\label{fig_3} 
Doping dependence of ${\bar \chi}_{\rm SS}$ (thick solid lines),
${\bar \chi}_{\rm CDW}$ (dashed lines),
and $\Delta_s/\omega_0$ (thin solid lines), for the Hol-Hub model (panels a and b),
and the Pei-Hub model (panels c and d), both with ${\bar U} = 0.1$ and $E_F/\omega_0 = 5$.
${\bar \chi}_{\rm SS}$ and ${\bar \chi}_{\rm CDW}$ were computed for $T/\omega_0 = 0.1$.
The values of $\lambda_{\rm Hol}$ and $\lambda_{\rm Pei}$
are labeled above each plot.
}
\end{figure}

Examining Fig. \ref{fig_1}d, we see that for moderate values of $\lambda_{\rm Pei}$,
for example near $\lambda_{\rm Pei} \approx 0.2$, $\chi_{\rm SS}$
is not divergent near $\delta = 0$, where $K_c^{\rm eff} < 1/2$,
nor is it divergent near $\delta = 1$, where $\Delta_s = 0$.  However, $\chi_{\rm SS}$ is
divergent for a certain range of moderate $\delta$.  Therefore, in these cases,
at fixed $T \ll \Delta_s$, $\chi_{\rm SS}$ exhibits a peak as a function of $\delta$ at an intermediate 
value of $\delta$.
This peak is shown explicitly in Figs. \ref{fig_2} and \ref{fig_3},
where we plot ${\bar \chi}_{\rm SS} \equiv \pi v_F \chi_{\rm SS}$
(thick solid line) versus $\delta$
at $T/\omega_0 = 0.1$, for representative parameters.
The CDW susceptibility
${\bar \chi}_{\rm CDW} \equiv \pi v_F \chi_{\rm CDW}$
(dashed lines) does not exhibit such a peak.

\begin{figure}
\includegraphics[width=0.485\textwidth]{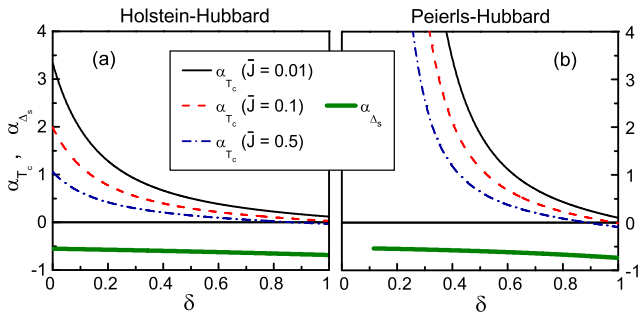}
\caption{\label{fig_4} 
Dependence of the isotope effect exponents $\alpha_{T_c}$ and
$\alpha_{\Delta_s}$ on the doping parameter $\delta$ and 
interchain coupling strength ${\bar J}$
for the Hol-Hub model with $\lambda_{\rm Hol} = 0.275$ (panel a) and
Pei-Hub model with $\lambda_{\rm Pei} = 0.25$ (panel b).
For both panels, ${\bar U} = 0.1$ and $E_F/\omega_0 = 5$.
$\alpha_{\Delta_s}$ is independent of ${\bar J}$.
}
\end{figure}

In Figs. \ref{fig_2} and \ref{fig_3} we also
plot $\Delta_s/\omega_0$ (thin solid lines), which shows that
at low dopings,
$\chi_{\rm SS}$ increases with increasing doping,
despite the fact that the superconducting pairing strength $\Delta_s$ decreases!
The reason for this discrepancy is the different doping dependencies of the effective interactions
in the charge and spin channels, which determine $K_c^{\rm eff}$ and $\Delta_s$ respectively.
It is worth mentioning that in the cuprates, the superconducting
gap also decreases with increasing doping, which
in the underdoped region occurs at the same time that
$T_c$ increases!

We now consider an array of weakly coupled quasi-1D chains
with dephased CDW, and
treat the interchain coupling $J$
on a mean-field level, which means that
$T_c$ is determined by the temperature at which $2 J \chi_{\rm SS} = 1$ \cite{erica,arrigoni}.
(The numerical prefactor 2 is determined by the number of nearest neighbor chains.) 
In this case, $T_c$ exhibits a peak at the same $\delta$
where $\chi_{\rm SS}$ is peaked
(assuming $J$ is doping independent). 
The isotope effect exponent $\alpha_{T_c}$ is readily
computed \cite{alpha}, and is shown versus $\delta$ in Fig. \ref{fig_4}
for various values of ${\bar J} \equiv J/(\pi v_F)$.
At low dopings, $\alpha_{T_c}$ is larger than the BCS value,
then drops below 1/2 as $\delta$ is increased.  Fig. \ref{fig_4}
also shows $\alpha_{\Delta_s}$, which is
weakly doping dependent and negative.
Note that there exists a range of $\delta$ for which
$|\alpha_{T_c}/\alpha_{\Delta_s}| \ll 1$.
Similarly, near optimal doping in the cuprates, 
$|\alpha_{T_c}/\alpha_{T^*}| \ll 1$,
where $\alpha_{T^*} = - d \ln T^*/ d \ln M < 0$, and $T^*$
is the pseudogap temperature \cite{pseudogap}. 

In some models \cite{erica} of high-temperature superconductivity
based on stripes \cite{stripes}, the concentration of holes on
a stripe remains fixed when the doping in the Cu-O plane changes,
but the spacing between the stripes changes.  In that case,
as the doping increases, the parameter $\delta$ remains fixed, but $J$ increases
due to the decreased spacing between stripes.  Then Fig. \ref{fig_4} predicts
that $\alpha_{T_c}$ again decreases with increasing doping.
In such a model,
in the underdoped region, where the stripes are far apart and represent
well defined quasi-1D chains, increasing the doping increases $T_c$
due to the increase in $J$.  But in the overdoped region, the stripes begin
to lose their 1D character.  This drives $T_c$ down since their quasi-1D
character was the reason for the high pairing scale.

To conclude,
in the interacting 1DEG,
the electron-phonon interaction can cause a strongly divergent superconducting
susceptibility with properties that are dramatically different from a Fermi liquid superconductor.
Using accurate analytic techniques, we
have studied microscopic models of quasi-1D electrons coupled
to phonons, and pointed out qualitative similarities
to the high-temperature superconductors.  These similarities include the doping dependence of $T_c$, 
the doping dependence of the superconducting pairing energy,
the doping dependence of the isotope effect on $T_c$,
and the sign of the isotope effect on the spin gap.

I would like to acknowledge useful discussions with S. Kivelson.
This work was supported by the Department of Energy contract 
No. DE-FG03-00ER45798.


\begin{references}
\bibitem{footnote1} In the MSRG technique, we approximate the phonon propagator as a step function
of frequency (see Ref. \protect\onlinecite{ian}).
\bibitem{alpha} We can use
$\alpha_{T_c}  = -\frac{1}{2}\frac{\Delta T_c/T_c} {\Delta \xi / \xi}$,
where $\xi \equiv E_F/\omega_0$, $\Delta \xi$ is infinitesimal,
$\Delta T_c \equiv T_c^\prime - T_c$, and
$T_c^\prime$ is the transition temperature after changing $l_0 \rightarrow l_0^\prime = \ln({\xi} + 
\Delta {\xi})$, $c \rightarrow c^\prime = c \, l_0/l_0^\prime$, and
$\delta \rightarrow 1 - c^\prime$.  The changes in $c$ and $\delta$ ensure that $E_F$ and $\mu$
remain fixed when changing $\xi \propto M^{1/2}$.
\end{references}
\end{document}